# Phonon-lithium ion interactions: A case study of LiM(SeO$_3$)$_2$

Runxin Ouyang[a], Yu Yang[a], Chaohong Guan[a], Hong Zhu[a,*]

a. *University of Michigan–Shanghai Jiao Tong University Joint Institute, Shanghai Jiao Tong University, 800 Dongchuan Road, Shanghai, 200240, China*

Corresponding author: * E-mail: hong.zhu@sjtu.edu.cn

**ABSTRACT**:

Li ion diffusion is fundamentally a thermally activated ion hopping process. Recently, soft lattice, anharmonic phonon and paddlewheel mechanism have been proposed to potentially benefit the ion transport, while the understanding of vibrational couplings of mobile ion and anions is still limited but essential. Herein, we access the ionic conductivity, the stability and the lattice dynamics in LiM(SeO$_3$)$_2$ (M =Al, Ga, In, Sc, Y, and La) with two types of oxygen anions within LiO$_4$ polyhedron, namely edge-shared and corner-shared, the prototype of which, LiGa(SeO$_3$)$_2$, has been experimentally synthesized. We studied in detail the anharmonic and harmonic phonon interactions, as well as couplings between vibrations of edge-bonded or corner-bonded anions in Li polyanions and Li ion diffusion. As M changing from Sc to La, anharmonic phonons increase alongside reduced activation energy for Li diffusion. Phonon modes involving edge-bonded oxygen anions contribute more to Li migration than corner-bonded oxygen anions, owing to greater atomic interactions between Li ions and edge-bonded anions. Thus, rather than the overall lattice softness, attentions shall be paid to reduce the frequency of the critical phonons contributing to Li ion diffusions as well as to increase the anharmonicity, for the design of Li ion superionic conductors for all-solid-state-batteries.

## 1. INTRODUCTION

The development of solid state electrolytes (SSEs)[1–4] with high ionic conductivities is critical for the all solid state batteries. In recent years, great progress has been made by cation and anion substitution in electrolyte materials for an optimal carrier concentration as well as a lower activation energy for Li or Na transport, such as lithium or sodium super-ionic conductors (LISICON[5] or NASICON)[6–8]. Such chemistry modulations are elucidated to often be accompanied with the softening of some vibrational modes relevant to Li diffusion. Prior work by Smith et al. demonstrated that substituting selenium for sulfur softens certain Li-P-S bending and stretching modes, which lowers Li migration barriers and increases Li diffusivity along 3D pathways in the β-Li$_3$PS$_4$[9] sulfide. Similar effects have been observed in the thiophosphate systems by substituting Si for P, namely Li$_{3.25}$Si$_{0.25}$P$_{0.75}$S$_4$.[10] Recent efforts by Park and colleagues to substitute Se for S in

Li$_6$PS$_5$I showed that enhanced Li mobility can be attributed to the softening of the PS$_4$ tetrahedra phonon modes.[13] Kim and coworkers demonstrated that materials with lower Li phonon frequencies are likely to exhibit lower activation energies below 300K, which have been observed in various electrolytes like LISICON, Li$_6$PS$_5$X (X=Cl, Br, I)[11–13] argyrodites, and NASICONs.[14–16]

In addition to soft phonons, anharmonic vibrations and phonon-ion coupling are also reported to play critical roles in enabling rapid ion diffusion.[17] In recent works, the anharmonicity of the crystals shows a great influence on the cation diffusions and thermal transports in Li-ion, Cu-ion[18], Ag-ion[19], and Na-ion[20] type superionic conductors. Anharmonic phonons in different phases of Na$_3$PS$_4$ have reluctantly frustrated the Na ion energetic migration pathways[17]. The anharmonic coupling of low-frequency Li phonon modes with high-frequency anion stretching or flexing phonon modes is reported to boost the Li ion conduction in a few representative sulfide and halide solid electrolytes, including Li$_7$P$_3$S$_{11}$, Li$_{10}$GeP$_2$S$_{12}$, β-Li$_3$PS$_4$, Li$_3$YBr$_6$ and Li$_3$ErCl$_6$[21]. Lately, Gordiz et al developed "Mass Diffusivity Modal Analysis" (MDMA) method to extract the subset of phonons that significantly contribute to the Li ion diffusivities, the excitation of which increased the Li ion diffusivities in Ge-substituted Li$_3$PO$_4$[22] by several orders of magnitude. Moreover, Kim et al. discovered that TiO$_6$ octahedral rocking modes contribute to the ion hopping in the Li$_{0.5}$La$_{0.5}$TiO$_3$ material and elucidated that the coupling of the TiO$_6$ octahedral rocking phonon and Li ion promotes the ion diffusion.[23]

Considering that ion transport is fundamentally a thermally activated ion hopping process, phonon-driven design approaches or phonon engineering show great promise for enabling next-generation SSEs with high conductivity, where a fundamental understanding of the critical phonon modes for Li diffusion as well as the effective way to tune it is in need. For example, the relationship between the local environment and phonon-ion migration coupling has not been systematically studied. So far, very few studies exist probing the local bonds in metal polyhedrons correlated to Li transport.[24] While progress has been made, open questions remain regarding how the phonon-ion interactions couple to ion mobility across different local environment, phonon mode and chemistry.

LiGa(SeO$_3$)$_2$, in which the Li polyhedron and the Ga polyhedron are connected through edge- or corner-shared oxygen atoms, has been screened out as an interesting lithium ion conductor with the similar structural features to NASICON[6] (i.e., regardless the Li polyhedron, the non-lithium cation polyhedra are corner shared by oxygen atoms) and later experimentally synthesized with the room temperature conductivity up to 0.11 mS cm$^{-1}$.[25] Compared to LiIn(IO$_3$)$_2$ and LiScAs$_2$O$_7$ with only edge-shared anions between the Li polyhedron and non-lithium polyhedron or Li$_{10}$GeP$_2$S$_{12}$ and Li$_3$PS$_4$ with only corner-shared anions, LiGa(SeO$_3$)$_2$ provides the opportunity to study dynamic couplings between lithium and differently bonded anions within the LiO$_4$ polyhedron. In this work, density functional theory (DFT) calculations have been conducted for the LiM(SeO$_3$)$_2$ (M =Al, Ga, In, Sc, Y and La) selenides to uncover the coupling of migrating Li and neighboring oxygen anions in different local environments. The harmonic and anharmonic phonon interactions in LiM(SeO$_3$)$_2$ are studied by MDMA methods, while the anharmonicities of the selenides are strongly inversely correlated with the energy barriers for Li diffusions. Moreover, by analyzing the phonon modes' contributions, it is found that the edge-bonded oxygen ($O_e$) vibrational phonon modes ($O_e$ modes) are more significantly coupled with Li ion diffusion compared to the corner-bonded oxygen ($O_c$) vibrational phonon modes ($O_c$ modes). Through the bonding analysis of the edge-bonded and corner-bonded oxygens, it is observed that the atomic interaction between Li and $O_e$ is larger than that between Li and $O_c$. Thus, the modulation of the anharmonic phonons, especially $O_e$ dominating ones, could enhance the Li ionic conductivities of the selenide-type superionic conductors. It appears important to investigate, soften and increase the anharmonicity of the critical phonon modes enabling lithium ion diffusion.

## 2. COMPUTATIONAL METHODOLOGY

All calculations were performed using the Vienna Ab-initio Simulation Package (VASP) software[26], based on the projector-augmented wave method within the framework of DFT. The generalized gradient approximation (GGA)[27] in the form of Perdew−Burke−Ernzerhof exchange[28] functional was used. To consider the strong correlation effects of the transition metal in $LiSc(GeO_3)_2$, $LiY(GeO_3)_2$ and $LiLa(GeO_3)_2$ selenides, both structural optimizations and electronic structure calculations were carried out using the spin-dependent GGA plus Hubbard correction U (GGA + U) method [29], with the effective $U_{eff}$ parameters of $5.00^{30}$, $3.00^{31}$, and $10.32^{32}$ eV for the Sc 3d, Y 4d, and La 4f states, respectively. The plane-wave energy cutoff was set to 520 eV. The Monkhorst−Pack method[33] with $2 \times 2 \times 2$ kpoint meshes was employed for the Brillouin zone sampling of $LiM(GeO_3)_2$ (M =Al, Ga, In, Sc, Y, and La) unit cells, respectively. The convergence criteria of energy and force were set to $10^{-5}$ eV/atom and 0.01 eV/Å, respectively. The chemical and electrochemical stability have been theoretically evaluated by the compositional phase diagram and lithium grand potential phase diagram, respectively, constructed by the Pymatgen code[34] based on the DFT ground-state energies of the corresponding compounds from the Materials Project database. Phonon calculations were performed by Phonopy and VASP with the higher convergence criteria of energy and force, $10^{-7}$ eV/atom and $10^{-5}$ eV/ Å, respectively. Activation energy and diffusion coefficients of Li were calculated based on Ab-initio molecular dynamics (AIMD) simulations for $2 \times 1 \times 1$ $LiM(SeO_3)_2$ supercells with the plane wave cutoff of 300 eV and the $1 \times 1 \times 1$ kpoint mesh grid. Time steps were set to 2 fs, and all the system were simulated for 20000 steps in the NVT statistical ensemble. The critical phonon modes for the Li ion diffusion were analyzed by Mass Diffusivity Modal Analysis (MDMA)[22] method. The crystal orbital bond index (COBI) analyses[35] were carried out for Li-O and M-O bonds in $LiM(GeO_3)_2$ by Lobster. The projected force constants (pFC) along the direction of A-B bond $\phi_p$ are calculated from harmonic force constants, generated as cartesian force constant matrices for each pair of A and B atoms as outlined in equation 1 through PHONOPY. [35] If a crystal structure is dynamically stable, matrix elements in equation 1 are calculated upon the potential energy change U with respect to the displacements of atom A and B (u) along the $\alpha$ and $\beta$ direction as defined in equation 2.

$$\phi_p(AB) = \left| \begin{pmatrix} \phi_{AB}^{xx} & \phi_{AB}^{xy} & \phi_{AB}^{xz} \\ \phi_{AB}^{yx} & \phi_{AB}^{yy} & \phi_{AB}^{yz} \\ \phi_{AB}^{zx} & \phi_{AB}^{zy} & \phi_{AB}^{zz} \end{pmatrix} \left\{ \frac{r(A)-r(B)}{|r(A)-r(B)|} \right\} \right| \quad \text{(Equation 1)}$$

$$\phi_{AB}^{\alpha\beta} = \frac{\partial^2 U}{\partial u_A^\alpha \partial u_B^\beta} \quad \text{(Equation 2)}$$

This method determines the force-constant for any pair of atoms within the supercell. The projected force constants along the Li ion migration direction (mpFC) can be calculated based on the equation 3, where $\vec{r_{Li}}$ are the Li ion migration directions.

$$m(\phi_p(AB)) = \phi_p(AB) \cdot \vec{r_{Li}} \quad \text{(Equation 3)}$$

To directly obtain the modes contributing to the ion diffusion from MD simulations, we used the definition of diffusivity based on the fluctuation-dissipation theorem and followed a modal decomposition approach similar to that used in MD-based thermal-transport studies.[36] In the "mass diffusivity modal analysis" (MDMA), the modal contributions to the atomic velocities are obtained and then used to determine the conductivity diffusivity as follows

$$D_\sigma = \frac{N_c}{3} \int_0^\infty \langle v_c(\tau) v_c(0) \rangle \, d\tau \quad \text{(Equation 4)}$$

Conductivity diffusivity ($D_\sigma$) not only includes the self-correlation information embedded in the tracer-diffusivity definition, but also includes the many-particle-correlation information. $N_c$ is the number of hopping particles in the system, $\langle \cdots \rangle$ in Equation 4 denotes the auto-correlation operator, and $v_c$ is the center of mass velocity for the hopping particles (as shown in Equation 5) and $v_i$ is the velocity of atom i.

$$v_c = \frac{1}{N_c} \sum_{i=1}^{N_c} v_i \quad \text{(Equation 5)}$$

The individual contribution by mode of vibration n to $v_i$ ($v_{i,n}$) can be determined by

$$v_{i,n} = \frac{\dot{Q}_n}{\sqrt{m_i}} e_{i,n} \quad \text{(Equation 6)}$$

where $e_{i,n}$ is the eigenvector for mode n, assigning the direction and displacement magnitude of atom i, $m_i$ is the mass of atom i and N is total number of the atoms. The modal velocity coordinate is $\dot{Q}_n = \sum_{i=1}^{N} \sqrt{m_i}\, e_{i,n}^* \cdot v_i$.

Based on Eq. 5 and 6, the hopping particles' center of mass velocity can be written as:

$$v_c = \sum_{n=1}^{3N} \frac{1}{N_c} \sum_{i=1}^{N_c} v_{i,n} \quad \text{(Equation 7)}$$

Combining Eq. 7 and 4, the conductivity diffusivity can be obtained by substituting the $v_c$ by $v_{i,n}$ as:

$$D_\sigma = \sum_{n=1}^{3N} \sum_{n'=1}^{3N} \frac{1}{3} \int_0^\infty \langle \sum_{i=1}^{N_c} v_{i,n}(\tau) \sum_{i'=1}^{N_c} v_{i',n'}(0) \rangle \, d\tau = \sum_{n=1}^{3N} \sum_{n'=1}^{3N} D_{\sigma,n,n'} \quad \text{(Equation 8)}$$

Individual contributions from the correlation/interaction of eigen mode pairs n and n' to conductivity, $D_{\sigma,n,n'}$, are given by,

$$D_{\sigma,n,n'} = \frac{1}{3}\int_0^\infty \langle \sum_{i=1}^{N_c} v_{i,n}(\tau) \sum_{i'=1}^{N_c} v_{i',n'}(0) \rangle d\tau \quad \text{(Equation 9)}$$

$D_{\sigma,n,n'}(n \neq n')$ presents the anharmonic phonon modes' contribution to the hopping diffusivity, while the value for n=n' represents the harmonic phonon modes' contribution.

## 3. RESULTS AND DISCUSSION

### 3.1. Stability and ionic conductivity.

The selenides LiM(SeO$_3$)$_2$ (M =Al, Ga, In, Sc, Y, and La) (in the I4$_2$d space group) contain the distorted Li tetrahedra (LiO$_4$), which are isolated from another LiO$_4$ by the trivalent metal octahedra (MO$_6$), as shown in the Figure 1a. Optimized lattice constants and atomic positions of LiM(SeO$_3$)$_2$ are listed in Table S1 and S2 in Supporting Information. The two oxygen atoms of LiO$_4$ are individually corner-shared with the adjacent MO$_6$ octahedra, named O$_c$ in Fig. 1b, while the other two O atoms of LiO$_4$ are edge-shared with one MO$_6$ octahedron, namely O$_e$. Such a connectivity between LiO$_4$ and MO$_6$ in LiM(SeO$_3$)$_2$ is named corner and edge bonding, respectively. It is worth noting that the other two O atoms in MO$_6$ which are not bonded to LiO$_4$ but SeO$_3$ are named O$_n$, here. The LiO$_4$ and MO$_6$ polyhedra could be connected only through corner O atoms, such as LiZnPO$_4$ (schematically shown in Figure. 1c),[37] or purely through edges of polyhedron, such as LiIn(IO$_3$)$_2$ (schematically shown in Figure. 1d).[25]

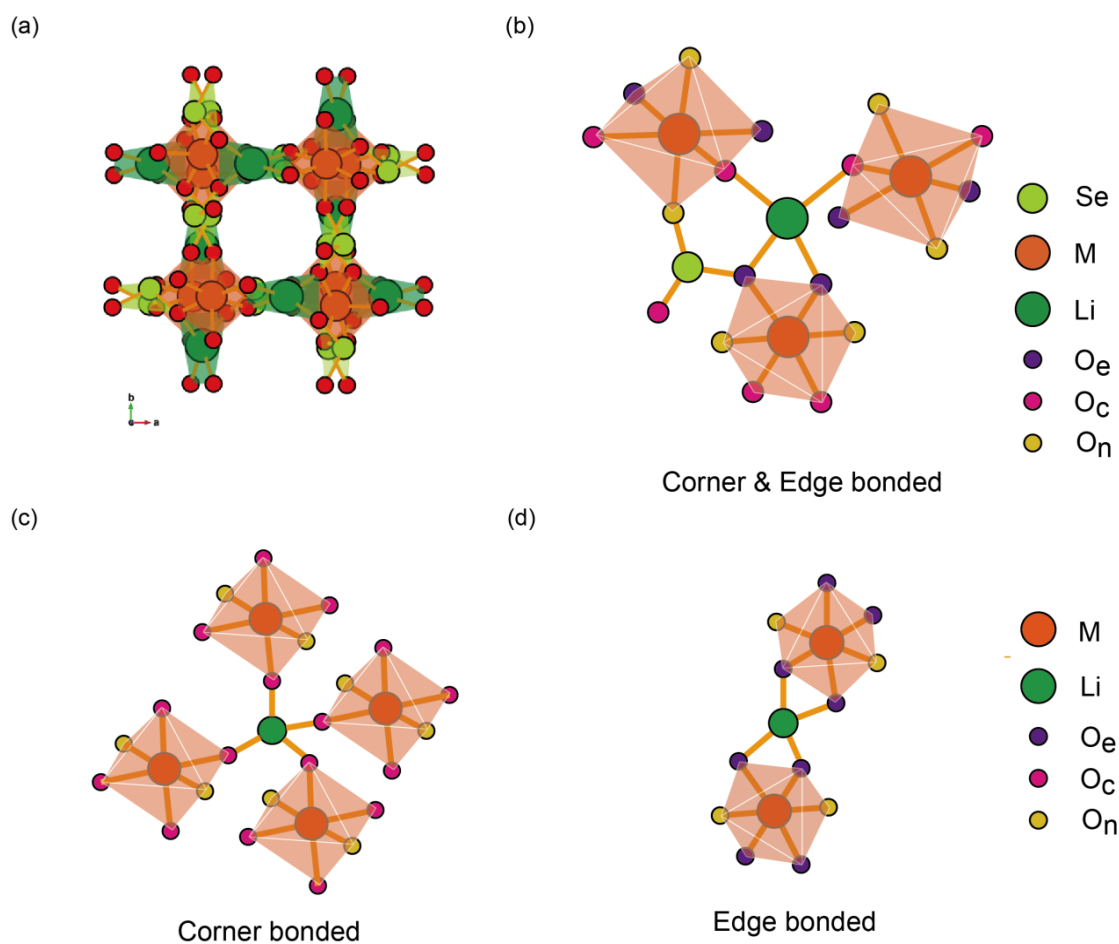

**Figure 1.** (a) Structure of LiM(SeO$_3$)$_2$ (M =Al, Ga, In, Sc, Y, and La) with the MO$_6$ octahedra and the distorted LiO$_4$ tetrahedra, where green, orange, lemon green and red spheres represent Li, M, Se and O atoms. (b) Connectivity between LiO$_4$ and MO$_6$ in LiM(SeO$_3$)$_2$ with the edge-bonded O atoms (O$_e$) in purple, the corner-bonded O atoms (O$_c$) in pink and non-lithium bonded O atoms (O$_n$) in yellow. Schematic plots of the other two connection ways between Li tetrahedron and M octahedra: (c) Corner bonded and (d) Edge bonded.

The LiM(SeO$_3$)$_2$ (M =Al, Ga, In, Sc, Y and La) selenides are thermodynamically stable based on the constructed Li$_2$O-M$_2$O$_3$-SeO$_2$ quaternary compound phase diagrams. The Li$_2$O-Sc$_2$O$_3$-SeO$_2$ phase diagram for LiSc(SeO$_3$)$_2$ is shown in Figure 2a, and similar plots for other LiM(SeO$_3$)$_2$ are elucidated in Figure S1. Based on the phonon band structures, all the LiM(SeO$_3$)$_2$ considered in this work are dynamically stable, as illustrated in Figure 2b and Figure S2. In addition, the elastic constants of LiM(SeO$_3$)$_2$ were calculated, as revealed in Table S3(SI), confirming the mechanical stabilities based on the Born criteria[41]. Therefore, in addition to LiGa(SeO$_3$)$_2$, LiM(SeO$_3$)$_2$ (M =Al, In, Sc, Y, and La) can be potentially synthesized experimentally based on their thermodynamic, dynamic and mechanical stabilities. Moreover, the elastic properties of SSE were crucial for the SSLIBs[38]. As proposed by Monroe et al, SSE materials with shear moduli larger than lithium metal can effectively suppress the growth of lithium dendrites.[39] From the comparisons in Table S3, the calculated Bulk (B), Shear (E), Young's (G) moduli of LiM(SeO$_3$)$_2$ are twice larger than those of Li metal, representing that LiM(SeO$_3$)$_2$ can restrain the growth of Lithium dendrites at their interface.

The calculated band gaps of LiM(SeO$_3$)$_2$ are in the range of 3.83-4.62 eV as depicted in Figure 2c. The element-projected band structures and density of states (Figure S3) reveal that valence band maxima are dominated by O$^{2-}$ anion, which are the first specie to be oxidized at high voltages, while the conduction band minima are dominated by M$^{3+}$ cation. During the cyclic operation of solid-state lithium ion batteries, the fairly high lithium chemical potential on the anode side induces the reduction reaction of the SSE material while the low lithium chemical potential on the charged cathode side causes the SSE material to oxidize and extract lithium.[40] The lithium grand potential phase diagram were built to appraise the electrochemical stabilities of LiM(SeO$_3$)$_2$ (see Figure 2a for LiSc(SeO$_3$)$_2$ and Figure S1 for others). The DFT calculated voltage profiles and phase equilibria of LiSc(SeO$_3$)$_2$ and LiM(SeO$_3$)$_2$ (M = Ga, In, Sc, Y, and La) upon lithiation and delithiation are illustrated in Figure 2c and Figure S4, respectively. The electrochemical window widths of LiM(SeO$_3$)$_2$ ranging from 0.48-2.36 V are much larger than that of sulfides, 0.10-0.75 V.[41] Thus, LiM(SeO$_3$)$_2$ selenides have much higher electrochemical stabilities and better compatibilities with electrodes than sulfides. It worthies to mention that the electrochemical stability analysis is carried out from the perspective of the thermodynamic driving force of the decomposition reaction. The kinetic factors, containing morphology, electronic and ionic of decomposition products, are important to the decomposition reaction, which are not considered here.

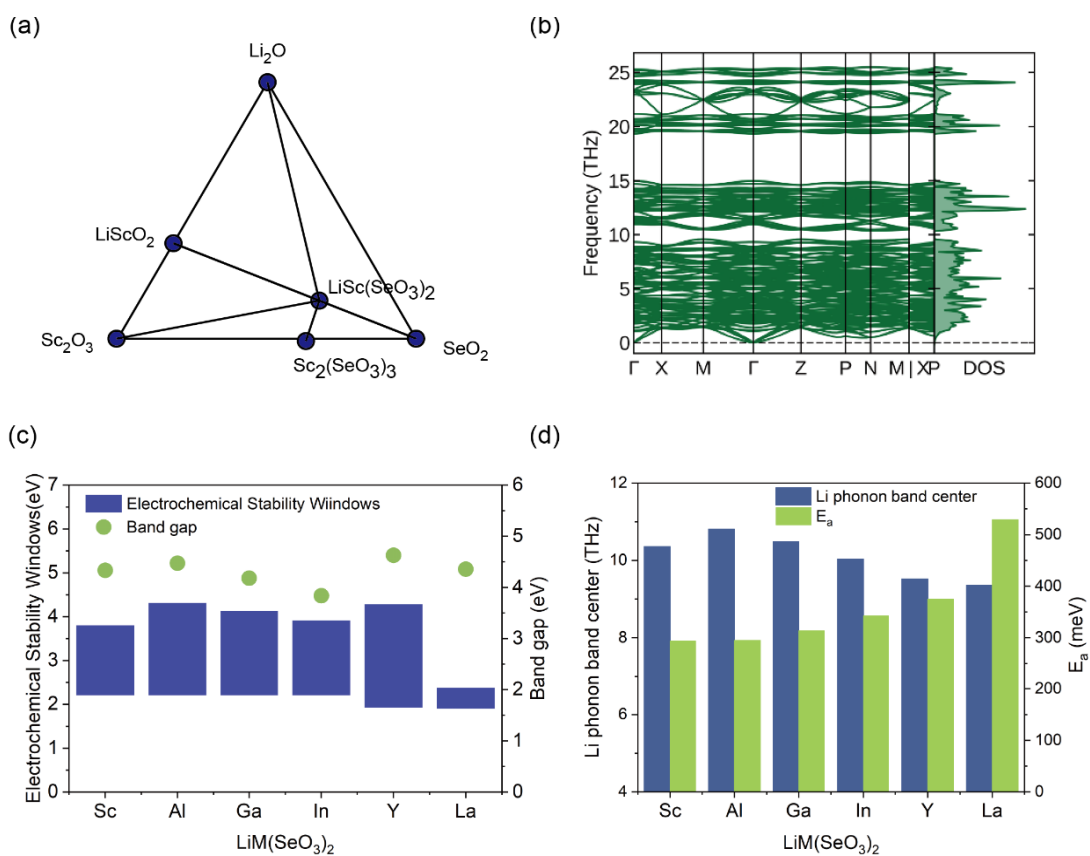

**Figure 2.** (a) Compound phase diagram of LiSc(SeO$_3$)$_2$ based on DFT calculation with blue points presenting the stable compounds. (b) Phonon dispersions and total phonon density of states (DOS)

of LiSc(SeO$_3$)$_2$. (c) Calculated electrochemical windows and band gap of LiM(SeO$_3$)$_2$. (d) Activation energy (E$_a$) for Li migrations of LiM(SeO$_3$)$_2$ determined by AIMD simulations and Li ion phonon band center of LiM(SeO$_3$)$_2$ determined by phonon DOS.

Based on the AIMD simulations, the activation energies for Li diffusion in LiM(SeO$_3$)$_2$ (M =Al, Ga, In, Sc, Y, and La) are 295.9, 314.3, 342.7, 294.4, 375.9, and 529.5 meV, respectively. Figure S5 show the Arrhenius plots of Li ion diffusion coefficients at different temperatures. Li ion diffusion in LiM(SeO$_3$)$_2$ is mainly based on the vacancy diffusion mechanism due to the homogenous and abundant vacancy sites. The extrapolated Li ion diffusion coefficient at 300K of LiM(SeO$_3$)$_2$ (M =Al, Ga, In, Sc, Y, and La) are 6.65×10$^{-9}$, 3.14×10$^{-9}$, 1.99×10$^{-9}$, 7.11×10$^{-9}$, 4.26×10$^{-11}$, 1.38×10$^{-11}$ cm$^2$/s, corresponding to Li ionic conductivities of 0.30, 0.13, 0.07, 0.27, 0.001, and 0.003 mS/cm. Moreover, the experimentally measured and DFT calculated Li ionic conductivity of LiGa(SeO$_3$)$_2$ are 0.11 mS/cm and 0.212 mS/cm respectively.[24] LiAl(SeO$_3$)$_2$ and LiSc(SeO$_3$)$_2$ seem to show even larger ionic conductivities at room temperature, compared to LiGa(SeO$_3$)$_2$.

### 3.2. Critical phonon modes for Lithium ion diffusion.

The Li projected phonon band center and the detailed phonon DOS as shown in Figure 2d and Figure S2, S6 indicate that LiM(SeO$_3$)$_2$ gets softer in the sequence of M=Al, Ga, Sc, In, Y, and La, which in general follows the same trend as the bulk moduli (see Figure S7) but show the opposite trend to the activation energies for Li diffusion. Such an observation that harder LiM(SeO$_3$)$_2$ has lower Li diffusion barrier is inconsistent with conventional understanding of the importance of soft lattice for superionic conductors.[15,16,42] Thus, besides the analysis of the overall lattice softness, it is critical to further investigate the correlation between different phonon modes and Li ion diffusion, especially those dominating Li transports. To better understand such a dynamic coupling, we performed the mass diffusivity modal analysis (MDMA, see methodology for details)[23,36] to quantify the anharmonic and harmonic phonons that facilitate or impede Li ion diffusion in LiM(SeO$_3$)$_2$. Based on 300K AIMD simulations and MDMA method, we calculated the modal contributions to Li ion diffusivity $D_{\sigma,n,n'}$ for the LiSc(SeO$_3$)$_2$ in Figure 3a and other LiM(SeO$_3$)$_2$ in Figure S8, which represent the contribution to Li ion diffusion from the correlation/interaction of phonon mode pairs for n and n' (see Equation 9). By summing over all the normalized $D_{\sigma,n,n'}$ for LiM(SeO$_3$)$_2$, we obtain the cumulative contributions to diffusivities from the correlation of phonons along the diagonal and off-diagonal terms indicating the harmonic and anharmonic phonon contribution (see Figure 3b). LiSc(SeO$_3$)$_2$, LiAl(SeO$_3$)$_2$ and LiGa(SeO$_3$)$_2$ with lower activation energy barriers display larger proportion of anharmonic contributions at low frequency as well as larger total anharmonic contributions, as illustrated in Figure 3b and S9a. Such an inverse relationship between the lithium ion migration activation energy barrier and the anharmonic contribution to lithium ion diffusivity in LiM(SeO$_3$)$_2$ is further verified by the other way to evaluate the anharmonicities (as shown in Figure S9b and S10), namely the difference of Li vibrational density of state (VDOS) changes in the range of low-frequency (N$_{low}$) and high-frequency (N$_{high}$) upon the cation-anion coupling, which have been used by Xu, et al for quantifying the anharmonicities of various SSEs.[21]

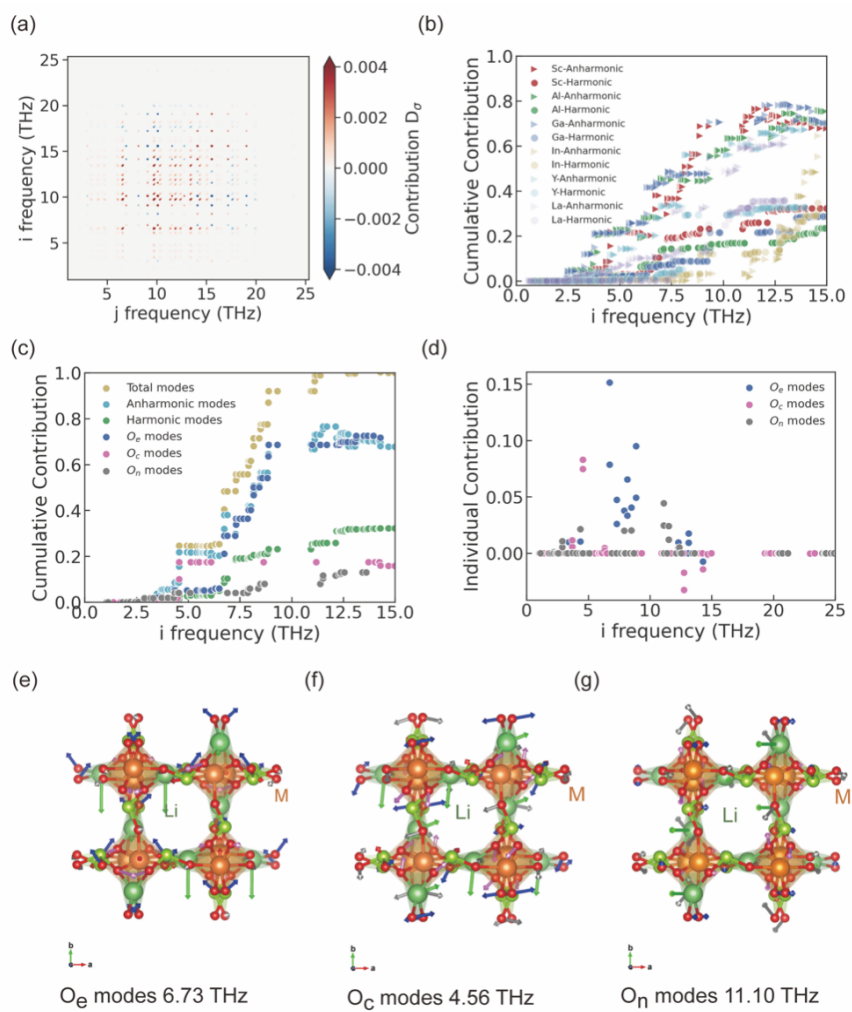

**Figure 3.** Modal contributions to Li ion migration and specific phonon mode with large contributions to Li diffusions. (a) Heatmaps for the magnitudes of the pairwise correlations/interactions of phonons contributing to the Li ion diffusivity of $LiSc(SeO_3)_2$ based on 300 K AIMD simulations; (b) The cumulative contributions of anharmonic (triangle) and harmonic (dot) phonon modes to the normalized energy required for ion hopping in $LiM(SeO_3)_2$ (M =Al, Ga, In, Sc, Y, and La); (c) The cumulative contributions of phonon modes ($O_e$ modes, $O_c$ modes, $O_n$ modes, anharmonic modes, harmonic modes and total modes are considered) to the normalized energy required for ion hopping in $LiSc(SeO_3)_2$; (d) The statistical plots of individual contribution of phonon mode (classified based on whether the largest vibration amplitude of O is $O_e$, $O_c$ or $O_n$) to the Li ion diffusion for $LiSc(SeO_3)_2$ at 300 K. Typical highly contributing phonon modes for Li ion diffusion of $LiSc(SeO_3)_2$: (e) $O_e$ phonon mode at 6.73 THz, (f) $O_c$ phonon mode at 4.56 THz, and (g) $O_n$ phonon mode at 11.1 THz (blue, pink and grey arrows represent vibrational vectors of $O_e$, $O_c$ and $O_n$, while green arrows represent vibrational vectors of Li).

Figure S9a elucidate that the proportion of anharmonic contributions to Li ion diffusion in $LiM(SeO_3)_2$ (M =Al, Ga, In, Sc, Y, and La) ranges from 52%-76%, while harmonic contributions are 23%-48%. $LiM(SeO_3)_2$ with the better ionic conductivity seems to have large anharmonic phonon contributions. In order to quantify coupling between the phonon modes and Li ion diffusion, we perform statistical analysis on the cumulative contributions of all phonon modes in LiSc(SeO$_3$)$_2$, for which the harmonic, anharmonic, O$_e$, O$_c$ and O$_n$ mode contributions to Li ion diffusions are illustrated in Figure 3c. The definition and calculation methods of phonon modes denoted as O$_e$ modes, O$_c$ modes and O$_n$ mode are defined based on which kind of O exibits the largest vibration amplitude for a given frequency (details are illustrated in supplement information). The cumulative contribution of the O$_e$ dominating modes to the Li ion migration is up to ~0.6 and larger than that of the O$_c$ or O$_n$ dominating modes in LiSc(SeO$_3$)$_2$ (other LiM(SeO$_3$)$_2$ show the similar trends). This demonstrates a stronger correlation between O$_e$ dominating phonon modes and Li ion diffusion than the other bonded O dominating phonon modes for LiM(SeO$_3$)$_2$. The individual contributions of O$_e$, O$_c$ and O$_n$ dominating phonon modes to Li ion diffusion of LiSc(SeO$_3$)$_2$ are shown in Figure 3d. It can be noticed that there are more O$_e$ dominating phonon modes contributing to Li diffusion, compared to the other two types of O dominating phonon modes. To better compare the phonon modes dominated by differently bonded O, we have specifically examined the most contributing vibration modes dominated by O$_e$, O$_c$, and O$_n$, which locate at 6.73 THz, 4.56 THz, and 11.1 THz and with a normalized contribution to Li diffusion of 0.15, 0.08 and 0.04, respectively, as shown in Figure 3e-g. It is not surprising to us that for the O$_e$ dominating mode at 6.73 THz, O$_e$ atom vibration shows larger amplitude than O$_c$ and O$_n$ atom vibrations, whose projection to Li diffusion direction is larger than the projection to the normal direction to Li diffusion. A detailed atom vibration amplitude and the projection information could be found in Table S4.

In order to better understand why the O$_e$ dominating modes have stronger correlation and contributions to Li ion diffusion than O$_c$ or O$_n$ modes, we specifically analyzed the local interaction of O$_e$-Li-O$_e$ and O$_c$-Li-O$_c$ bonds of LiO$_4$ polyhedron in LiM(SeO$_3$)$_2$ (M =Al, Ga, In, Sc, Y, and La) by integrated crystal orbital bond index (ICOBI) in Figure 4a. The -ICOBI values for O$_e$-Li-O$_e$ are higher than that for O$_c$-Li-O$_c$, indicating that interaction between Li and edge-bonded oxygen is stronger than that between Li and corner-bonded oxygen. This implies that vibration of edge-bonded O$_e$ anions with strong bonding strength with Li ions enables greater dynamic coupling with Li diffusion than corner-bonded O$_c$ anions, where similar ICOBI results for other systems are depicted in Figure S12 and S13. For the further validation, we also demonstrate the larger contribution of edge-bonded O over corner-bonded O to Li ion diffusion by performing the 900K AIMD simulation with fixed edged or cornered anion in LiAl(SeO$_3$)$_2$ and LiGa(SeO$_3$)$_2$, where Li ion diffusivities with fixed edged anion are lower than that with fixed cornered anion as depicted in Figure S13, S14.

To probe the correlation between local bonding strength and Li ion diffusion, the projected force constants along the Li ion migration direction (mpFCs) are applied for evaluating the contribution of the local atomic bonding to Li ion migration. The larger mpFC represents the larger atomic force interaction along the direction of Li ion migration. We elucidate that the mpFCs of O$_e$-Li-O$_e$ bonds are larger than the O$_c$-Li-O$_c$ bonds in LiM(SeO$_3$)$_2$ as illustrated in Figure 4b. It is concluded that O$_e$-Li-O$_e$ provides larger atomic force interaction along the direction of Li ion migration than O$_c$-Li-O$_c$. Moreover, the correlation between E$_a$ and the difference of mpFCs between the O$_e$-Li-O$_e$ and O$_c$-Li-O$_c$ bonds ($\Delta pFC = \text{mpFC}_{O_e-Li-O_e} - \text{mpFC}_{O_c-Li-O_c}$) in Figure 4c show that larger $\Delta pFC$ impose lower E$_a$ in LiM(SeO$_3$)$_2$, which could lead to asymmetric bonds within the LiO$_4$ polyhedron and hence higher anharmonicity. At last, considering that the $O_e$ dominating modes rather than the whole phonon spectrum are highly related to Li ion diffusion, we have studied the relationship between the $E_a$ for Li ion diffusion and the average frequency for $O_e$ dominating phonon modes calculated from 300k AIMD simulation for $LiM(SeO_3)_2$ (see Figure 4d). The lower average $O_e$ mode frequency in the $LiM(SeO_3)_2$ selenides is clearly accompanied with a lower activation energy barrier for Li diffusion. We have also investigated the phonon dynamics and Li diffusion for other Lithium oxide electrolytes with pure $O_e$ or $O_c$ anions, as shown in Table S5. It can be noted that, if the system only has one type of oxygen anion, the anion dominating phonon band center follows the similar trend to Li phonon band center as well as the Li diffusion activation energy. If the material contains differently bonded anions, the average frequency of edge-bonded oxygen shows a positive correlation with the Li diffusion barrier but an inverse relationship with the Li phonon band center. Thus, it seems to be essential to study the critical phonon mode as well as the local atomic interactions rather than the overall lattice softness for the design of superionic conductors, especially those with more complicated local anion bonding environments, such as $LiM(SeO_3)_2$ with the coexistence of corner-shared and edge-shared oxygen anions.

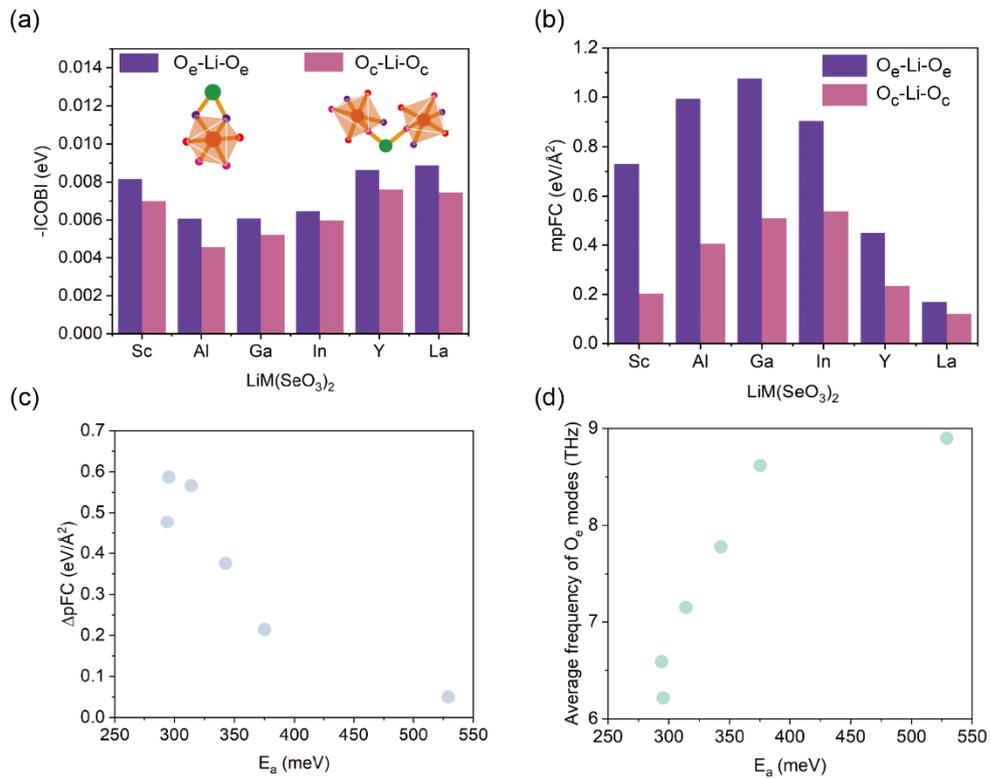

**Figure 4.** (a) The integrated crystal orbital Bond index (ICOBI) of $O_e$-Li-$O_e$ and $O_c$-Li-$O_c$ bonds for $LiM(SeO_3)_2$ (M =Al, Ga, In, Sc, Y, and La). (b) Projected force constants between $O_e$-Li-$O_e$ and $O_c$-Li-$O_c$ bonds along the Li migration directions (mpFC) for $LiM(SeO_3)_2$ (M =Al, Ga, In, Sc, Y, and La). (c) The correlation between $E_a$ and the difference of projected force constants ($\Delta pFC$) between $O_e$-Li-$O_e$ and $O_c$-Li-$O_c$. (d) The average frequency of $O_e$ dominating modes to Li ion diffusion vs. the DFT-calculated activation energy barriers in selenides.

## 4. CONCLUSIONS

In this work, the selenides electrolytes LiM(SeO$_3$)$_2$ (M =Al, Ga, In, Sc, Y, and La) have been theoretically explored in terms of lithium ionic conductivity, deformability, and chemical and electrochemical stability. Our DFT calculations show these selenides have promising properties as solid electrolytes. This combination of facile processability, wide electrochemical window, and fast ion transport makes these selenides attractive for further exploration. The LiM(SeO$_3$)$_2$ with hard overall lattice dynamics shows lower Li diffusion barrier, inconsistent with conventional understanding of the importance of soft lattice for superionic conductors.[15,16,42] Besides the analysis of the overall lattice softness, it is critical to further investigate the correlation between different phonon modes and Li ion diffusion, especially those dominating Li transports. Both harmonic and anharmonic phonon interactions in LiM(SeO$_3$)$_2$ are studied using the Mass Diffusivity Modal Analysis (MDMA) method. Results show a strong inverse correlation between anharmonicity and activation energy barriers in these selenides. Analyzing individual phonon mode contributions reveals more significant coupling between Li diffusion and vibrations of edge-bonded oxygen (O$_e$) versus corner-bonded oxygen (O$_c$) between LiO$_4$ polyhedron and MO$_6$ polyhedron, which has been validated by the stronger bonding strength analyses for O$_e$-Li-O$_e$ than that for O$_c$-Li-O$_c$. Thus, rather than the overall lattice softness, the reduced frequency of the critical phonons for Li diffusion, namely average O$_e$ mode for LiM(SeO$_3$)$_2$ selenides, facilitates the Li ion diffusion. Modulating anharmonic phonons, especially O$_e$-dominated ones, may enhance the asymmetry of the LiO$_4$ polyhedron as well as Li ionic conductivity in these selenide superionic conductors. To improve the Li ion diffusivity, it seems critical to study the critical phonon mode for Li ion diffusion as well as the local atomic interactions, from which higher anharmonicity through asymmetric local environments and lower frequency for the phonons contributing to Li diffusion may be better to be realized for higher ionic conductivity.

## ASSOCIATED CONTENT

**Supporting Information**

## DATA AVAILABILITY

The Data analyzed and calculated in this work are openly available at the websites of https://github.com/runxin123/Selenides/.

## CODE AVAILABILITY

The Data for buliding MDMA model and extract harmonic phonons codes are publicly available at Github website, https://github.com/runxin123/Selenides/.

## AUTHOR CONTRIBUTIONS

H.Z. designed the research and wrote the paper, R.O. performed the DFT calculations and analyzed the data; Y.Y. and C.G. revised the paper and supply suggestions. All authors discussed and commented on the paper.

**COMPETING INTERESTS**

The authors declare no competing financial interest.

**AUTHOR INFORMATION**

**Corresponding Author**

*E-mail: [1].

ORCID Hong Zhu: 0000-0001-7919-5661

**ACKNOWLEDGMENTS**

This work was supported by National Natural Science Foundation of China (52072240), and the Materials Genome Initiative Center at Shanghai Jiao Tong University. All simulations were carried out with computational resources from Shanghai Jiao Tong University High Performance Computing Center.